      [1999/12/01 v1.4c Il Nuovo Cimento]
\documentclass[varenna]{cimento}
\newcommand{\be}{\begin{equation}}
\newcommand{\ee}{\end{equation}}
\newcommand{\bea}{\begin{eqnarray}}
\newcommand{\eea}{\end{eqnarray}}

\usepackage{graphicx}
\title{Fermi-Bose mixture with tunable interactions}
\author{Giovanni Modugno}

\institute{LENS and Dipartimento di Fisica, Universit\`a di
Firenze\\  Via Nello Carrara 1, 50019 Sesto Fiorentino, Italy}

\begin{document}

\maketitle

\section{Introduction}
The possibility of controlling the atom-atom interaction in
ultracold quantum gases has recently opened various research
directions, ranging from molecular physics in the nanoKelvin regime
\cite{molecules} to strongly interacting superfluid Fermi gases
\cite{fermi} and to strongly correlated atomic systems
\cite{strongly}. The interaction is usually controlled through
Feshbach resonances \cite{inouye}, which consist in a resonant
coupling of atomic and molecular levels that can be tuned through an
external magnetic field.

This kind of tool has been extensively applied so far only to
homonuclear Bose and Fermi systems. Tuning of the interactions in
heteronuclear mixtures via Feshbach resonances can give access to an
even broader range of phenomena. Interaction effects such as
collapse or phase separation \cite{molmer} would be accurately
studied. A tunable interaction between the fermionic and bosonic
components would also give access to a wealth of condensed-matter
phenomena in strongly correlated systems that have been recently
proposed \cite{novelqp}. Moreover, Feshbach resonances can be used
to associate ultracold heteronuclear molecules. These would possess
a permanent electric dipole moment, allowing to manipulate the
samples via electric fields, and to study dipole-dipole interactions
in the context of ultracold quantum gases. The presence of a
long-range anisotropic interaction between the particles in Bose and
Fermi quantum gases is expected to profoundly modify the properties
of such systems, and to give access to new phenomenology
\cite{fmgases,zoller}.

The study of Feshbach resonances in heteronuclear systems has been
delayed by a few years with respect to the homonuclear case, mainly
because of the larger complexity of the experiments and because of
the lack of accurate theoretical models for Feshbach resonances in
most of the alkali pairs. Currently, Feshbach resonances have been
experimentally detected in three alkali fermion-boson mixtures:
$^{6}$Li-$^{23}$Na \cite{mit}, $^{6}$Li-$^{7}$Li \cite{ens},
$^{40}$K-$^{87}$Rb \cite{jila,ferlaino}. However, a fine tuning of
the interaction has so far been achieved only in the
$^{40}$K-$^{87}$Rb system \cite{zaccanti,ospelkausF}.

In this work we first describe the various steps that have led to
the observation of Feshbach resonances in the K-Rb system and their
accurate characterization. We then describe recent experiments in
which Feshbach resonances are exploited to study interaction effects
and to perform first experiments on association of weakly bound KRb
dimers.

\section{Feshbach resonances in the K-Rb mixture}

Among ultracold Fermi-Bose atomic mixtures, the $^{40}$K-$^{87}$Rb
system has so far been the most investigated one. Since its first
production \cite{roati}, it has been employed for a series of
experiments on interaction phenomena and on Fermi and Bose gases in
optical lattices. Nowadays it is by far the most employed ultracold
atomic mixture. One of the peculiarities of this system that made it
particularly interesting from the very beginning is the naturally
attractive interaction between fermions and bosons. This is
interesting to study mean-field effects such as collapse and bright
solitons or boson-induced superfluidity. The possibility of tuning
of the interspecies interaction via Feshbach resonances will further
enlarge the spectrum of phenomena that can be investigated with this
mixture.

Let us remind here a few general aspects of interactions in
ultracold atomic gases and Feshbach resonances.  The two-body
interaction between atoms is described by a contact potential and
the collisional events can be expanded in partial waves. At the low
temperature of actual experiments, only the lowest order,
energy-independent partial wave, i.e. the $s$-wave, need to be taken
into account. Just one parameter, the $s$-wave scattering length
$a$, is needed to define quantities such as the collisional
cross-section and the coupling constant for interacting quantum
gases in the mean-field approach. In a mixture of spin polarized
fermions and bosons, two different interactions have to be
considered: the fermion-boson interaction, described by $a_{FB}$,
and the boson-boson one, described by $a_{BB}$. Fermion-fermion
interaction must not be taken into account, since Pauli principle
strictly forbids $s$-wave interaction between identical fermions.
Throughout this work we consider the case in which only the
interspecies scattering length $a_{FB}$ can be tuned, while $a_{BB}$
is fixed to its natural value (this is about 100 $a_0$ for
$^{87}$Rb). In general, the possibility of tuning at the same time
both $a_{FB}$ and $a_{BB}$ in a fermion-boson system is marginal.
The scattering length $a_{FB}$ depends on the phase shift that the
wavefunction of a pair of free atoms experiences when the pair
separation approaches those typical of the molecular potential, i.e
several tens of $a_0$. As a matter of fact, the natural value of
$a_{FB}$ is determined just by the binding energy of the last
vibrational state of the molecular potential. For example, in the
case of the $^{40}$K-$^{87}$Rb pair in its ground state $a_{FB}$ is
negative and relatively large ($a_{FB}$=-185 $a_{0}$) because the
closest molecular level to the atomic threshold is actually just
above the threshold. By applying an external magnetic field to the
system, it is possible to tune in a different way the energy of the
atomic threshold and that of the molecular states, and eventually
obtain a situation in which a molecular level crosses the atomic
threshold. At this crossing the scattering length of the system
shows a resonant enhancement, the so called Feshbach resonance, and
passes from -$\infty$ to +$\infty$ as the molecular state passes
from above to below the atomic threshold. At one of such resonances,
the magnetic field dependence of $a_{FB}$ can be parametrized as
\begin{equation}
a_{FB}(B)=a_{bg}(1-\Delta/(B-B_0))\,,
\end{equation}
where $a_{bg}$ is the background value of the scattering length,
$B_0$ is the magnetic-field location of the resonance center, and
$\Delta$ is a width parameter that accounts for the coupling
strength of the specific atomic and molecular levels.

The $^{40}$K-$^{87}$Rb has a large number of internal states (the
total angular momentum is F=9/2 for K and F=1 for Rb in their ground
hyperfine state) and therefore in principle a large number of
Feshbach resonances is available. In a series of experiments on cold
collision measurements \cite{ferrari,sloshing,simoni} we have
determined to a good accuracy the scattering length $a_{FB}$ of the
system, which has been used to fine tune the best approximation of
the K-Rb molecular potential, and in particular to determine the
binding energy of the states close to dissociation. Actually, two
different molecular potentials appear for alkali atoms, i.e. the
singlet $X^1\Sigma^+$ and triplet $a^3\Sigma^+$ potentials with
scattering lengths $a_s$ and $a_t$, respectively. This study has
then allowed to predict the magnetic-field position of Feshbach
resonances \cite{simoni}. In particular, for most of the states we
expected several resonances in the region between 400 and 700 G.
This prediction was confirmed by a first study of heteronuclear
resonances in the absolute ground states performed at JILA
\cite{jila}. In a successive detailed investigation of Feshbach
resonances in various internal states \cite{ferlaino} we got the
complete picture of such resonances that is necessary for their
exploitation in experiment. In the next section we describe the
experimental and theoretical methods used.

\section{Feshbach spectroscopy}
In our experiment the K-Rb Fermi-Bose mixture is routinely produced
via laser and evaporative cooling in a magnetic trap
\cite{science,roati}. In order to study magnetic Feshbach resonances
one has to decouple the trapping mechanism from the magnetic moment
of the atoms. We achieved this by transferring the atoms from the
magnetic trap to an optical dipole trap created by focused laser
beams crossing in the horizontal plane. In a first series of
experiments \cite{ferlaino} we have used a titanium:sapphire laser
at a wavelength of 830~nm, which is sufficiently detuned from the
atomic resonance to provide a lifetime of the atomic sample
exceeding 1~s. The trap depth was about 10~$\mu$K, which is
sufficiently large to hold "hot" samples at temperatures above
quantum degeneracy. Typically 10$^5$ K fermions and 5$\times$10$^5$
Rb bosons were cooled in a magnetic trap down to temperatures of a
few hundreds nK and then adiabatically transferred to the optical
trap. The typical density and temperature of the bosonic sample in
the optical trap throughout this experiment were
5$\times$10$^{12}$~cm$^{-3}$ and 1~$\mu$K, respectively. The
fermionic sample was in thermal equilibrium with the bosonic one.

The mixture was initially prepared in the state $|F^K=9/2,
m^K_F=9/2\rangle \otimes |F^{Rb}=2, m^{Rb}_F=2\rangle$, which is the
only stable combination of low-field seeking states. This state does
not possess Feshbach resonances, since it has the maximum projection
of the angular momentum. In the choice of a K$\otimes$Rb state to
explore for Feshbach resonances one has to take into account its
stability against spin-exchange collisions. These are inelastic
collisions which can take place every time a second state with the
same projection of the total angular momentum $m^K_F$+$m^{Rb}_F$
with a lower energy is available. The excess kinetic energy is
tipically enough to have an immediate loss of the collision
partners. Due to the particular level structure of this K-Rb
mixture, all the states where both species are in their ground
hyperfine state and either Rb or K are in their absolute ground
state are stable against such collisions and have a lifetime that is
limited mainly by three-body recombination. According to our
theoretical study \cite{simoni}, most of these states were also
expected to feature Feshbach resonances.

To prepare the atoms in these states we have used a series of
radio-frequency (RF) and microwave ($\mu$w) adiabatic rapid
passages.  A first combination of sweeps was used to transfer the
system to its ground state $|9/2, -9/2\rangle \otimes |1, 1\rangle$
in presence of a bias field of about 10~G.  The field was then
raised to 100~G to perform the additional transfer of K or Rb to
excited states such as the $|9/2, -7/2\rangle$, $|9/2, 7/2\rangle$
or $|1, 0\rangle$.
\begin{figure}[tbp]
\includegraphics[width=\columnwidth,clip]{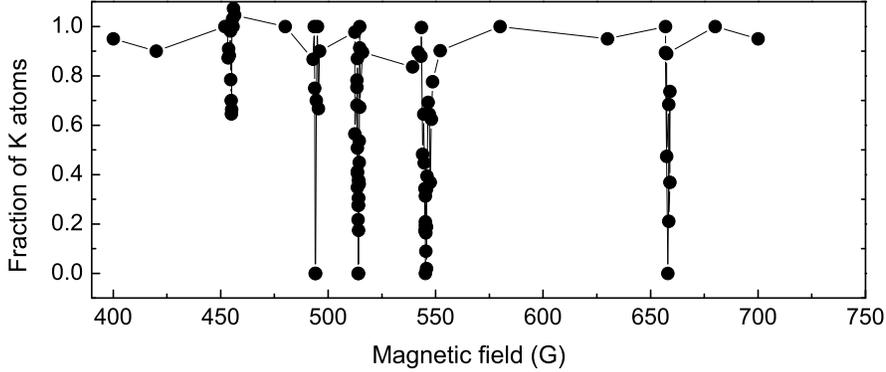}
\caption{Relative inelastic losses of potassium atoms in a
$^{40}$K-$^{87}$Rb mixture in its absolute ground state at
interspecies Feshbach resonances. The two features near 456~G and
515~G are $p$-wave resonances, the others are $s$-wave resonances.}
\label{fig1}
\end{figure}

The homogeneous field was then changed in few ms to any value in the
range 0-1000~G and actively stabilized there with an accuracy of
about 200~mG. The field was calibrated by means of RF spectroscopy
of the $|2, 2\rangle$ - $|2, 1\rangle$ transition of Rb.

As we will describe in detail in the next section, a Feshbach
resonance is usually accompanied by an enhancement of the inelastic
decay rate. The easiest way to look for Feshbach resonances is
therefore to record the fraction of atoms lost from the trap when
changing the magnetic field. In our mixture the best sensitivity to
resonances was achieved by recording the atom number of the minority
component, i.e. potassium, after about 1~s of permanence at a fixed
magnetic field. The typical experimental signature of interspecies
resonance is shown in Fig.~\ref{fig1} for the absolute ground state
of the mixture: due to the lower abundance of K in our sample, we
usually observe a complete loss of K atoms at resonance. To avoid
any possible confusion with homonuclear resonances we also check for
the absence of losses after removing either K or Rb from the mixture
before applying the magnetic field. Actually, in our investigation
we have observed several Rb resonances that have been previously
studied in dedicated experiments \cite{munich}.

In a first survey we have detected 14 Feshbach resonances in three
different states, and used them to refine our theoretical quantum
collisional model. To do this, the singlet $X^1\Sigma^+$ and triplet
$a^3\Sigma^+$ interaction potentials were parameterized in terms of
the $a_s$ and $a_t$ scattering lengths, respectively. The two-body
elastic cross section was then computed for different $a_{s,t}$ and
its maxima, corresponding to Feshbach resonances, were compared to
four sample experimental features. Once a good agreement was found,
a least square fit to all 14 experimental features was then
performed, letting also the van der Waals coefficient $C_6$ free to
vary. The best fit parameters with one standard deviation we
obtained are $a_s=(-111\pm5)a_0$, $a_t=(-215\pm10)a_0$ and
$C_6=(4292\pm 19)$~a.u., which agrees to better than one standard
deviation with the high precision {\it ab-initio} calculation
obtained in \cite{derevianko}. Error bars also include a typical
$\pm 10\%$ uncertainty in $C_8$. The position and width of the
resonances are listed in Table \ref{table1}. The average
theory-experiment deviation for the resonance positions is about
0.3~G only.

A determination of the K-Rb scattering lengths based on extensive
Feshbach spectroscopy is by far the most precise one can obtain. A
large number of detected resonances indeed unequivocally maps the
last few molecular states close to dissociation, and therefore
determines the scattering lengths. Our determination of $a_{s,t}$
helped in solving some existing inconsistencies between previous
experiments on cold collisions \cite{sloshing,goldwin}, collapse
\cite{collapse,hamburg} and Feshbach resonances \cite{jila} in this
mixture. The optimized $^{40}$K-$^{87}$Rb model can also be used to
determine singlet and triplet scattering lengths for any K-Rb
isotopic pair \cite{ferlaino}. Within the Born-Oppenheimer
approximation this can be simply achieved by using the appropriate
reduced mass in the Hamiltonian. Such mass-scaling procedure depends
in a sensitive way on the actual number of bound states supported by
the potentials. They are nominally $N_b^{s}=98$ and $N_b^{t}=32$ for
the singlet and triplet {\it ab initio} potentials we use, with an
expected uncertainty of $\pm 2$~\cite{zemke}. Other quantities of
general interest for experiments with K-Rb mixtures are the
effective elastic scattering length $a$ and the location of Feshbach
resonances for the absolute ground state. We have determined also
these quantities with relatively high accuracy using the collisional
model above; they can be found in Ref.~\cite{ferlaino}. These data
complete our previous investigation \cite{ferrari}, and are of great
interest for forthcoming experiments on Bose-Bose mixtures with K
and Rb atoms \cite{modugno,minardi}.

\begin{table}[t]
\begin{center}
\caption{Magnetic-field positions and widths of the observed
$^{40}$K-$^{87}$Rb resonances compared to the corresponding
theoretical predictions of our best-fit model. $\Delta_{\rm th}$ is
defined in the text, and $\ell$ is the orbital angular momentum of
the molecular state associated to each resonance.} \label{table1}
\vskip 12pt
\begin{tabular}{l | c c c cc}
\hline \hline
 $|m_{fa}\rangle\otimes|m_{fb}\rangle$& $B_{\rm exp}$ (G) &  $B_{\rm th}$(G)&$-\Delta_{\rm th}$ (G)& $\ell$ \\
 \hline \hline
% \parbox[l]{5cm}{$|-9/2\rangle$+$|1\rangle$}  &--&--&200&xx&0\\
$|-9/2\rangle\otimes|1\rangle$
& 456.0 &  456.5 &$2~10^{-3}$&1\\
& -- &  462.2   & 0.067  &0\\
&495.6 & 495.7 &0.16 &0\\
&515.7 &515.4 & 0.25  &1\\
&546.7&546.8&2.9&0\\
&547.4&547.6&0.08&2\\
&658.9&659.2& 1.0 &0\\
&663.7&663.9& 0.018 &2\\
\hline
$|-7/2\rangle\otimes|1\rangle$& 469.2&469.2&0.27&0\\
&--&521.6&0.051&0 \\
&584.0&584.1&0.67&0 \\
& 591.0 &591.0 &$2~10^{-3}$&2 \\
& 598.3 & 598.2&2.5&0\\
& 697.3 &697.3 &0.16&0 \\
& 705.0 &704.5 &0.78&0\\
\hline $|7/2\rangle\otimes|1\rangle$&299.1&298.6&0.59&0\\
& 852.4 &  852.1&0.065& 0\\
\hline \hline

\end{tabular}
\end{center}
\end{table}

Let us now discuss the properties of the Feshbach resonances we
detect in the experiment, listed in Tab.~\ref{table1}. The nature of
the molecular states associated to the resonances can be better
understood through multichannel bound state calculations. Since
$a_s$ and $a_t$ are comparable the spacing between singlet and
triplet vibrational levels is small compared to the hyperfine
interaction.  Strong singlet/triplet mixing then occurs at least for
the two vibrational states closest to dissociation, resulting in
molecular levels labeled as $(F^K F^{Rb} F \ell)$ in {\it zero}
magnetic field, where $\ell$ is the rotational quantum number and
$\vec F = \vec{F}^K +\vec{F}^{Rb}$. The features below
$\approx600$~G arise from these strongly mixed levels. At such
magnetic fields however the Zeeman magnetic energy is comparable to
the smaller hyperfine splitting in the system, that of $^{40}$K.
Therefore $F^K$ is not a good quantum number to label the
resonances, whereas $F^{Rb}$ is approximately good and equal to 2.
Resonances at higher magnetic field correlate with more deeply-bound
states and tend to assume singlet or triplet character. In all cases
$\ell$ is an almost exact quantum number and is also shown in
Tab.~\ref{table1}.

The measurement of inelastic decay features described above allows
to access only the location of Feshbach resonance, but not the exact
magnetic field dependence of the fermion-boson scattering length.
This can be calculated by our collisional model: isolated resonances
are well described by the simple parametrization introduced above:
$a_{FB}(B)=a_{bg}(1-\Delta_{th}/(B-B_0))$. The background scattering
length for a particular state is determined by the model with an
accuracy comparable to that obtained for $a_{t,s}$. For example, the
determined value for the ground state is
$a_{bg}$=(-185$\pm$7)~$a_0$. The width $\Delta_{th}$ can also be
calculated for all resonances, and is shown in Tab.~\ref{table1},
together with the angular momentum $\ell$ of the corresponding
molecular states. The $\ell$=0 molecules tend to give rise to broad
resonances due to strong spin-exchange coupling to incoming $s$-wave
atoms. We also observe a few narrow resonances due to coupling of a
$\ell$=2 molecule to incoming $s$-wave atoms through weaker
anisotropic spin-spin interactions~\cite{cs,stuttgart}. The two
resonances associated with $\ell$=1 molecules couple by
spin-exchange to incoming $p$-wave atoms. These resonances have an
energy-dependent width \cite{bohnp} which we compute at collision
energy $E/k_B \approx 1 \mu$K, with $k_B$ the Boltzmann constant.
Tab.~\ref{table1} also shows two narrow not yet observed resonances.
We find that several stable states of the mixture present at least
one broad resonance, analogous to that in the ground state near
546~G. Any of these resonances can be very well suited for control
of the interaction and molecule formation.

\begin{figure}[htbp]
\includegraphics[width=\columnwidth,clip]{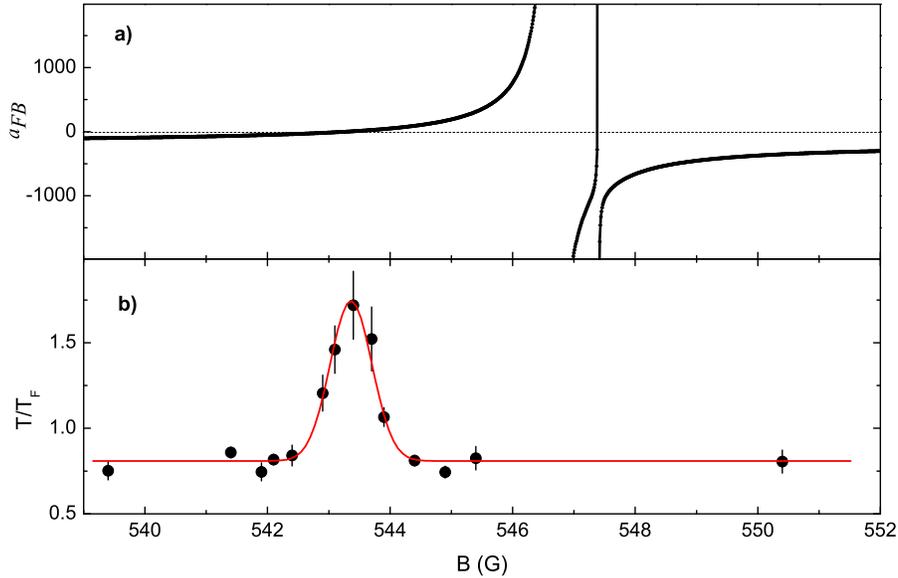}
\caption{a) Theoretical expectation for the fermion-boson scattering
length $a_{FB}$ at the broadest Feshbach resonance in the ground
state of the K-Rb mixture. b) Efficiency of sympathetic cooling of
fermions around the zero-crossing location. The feature at 547.4~G
is a narrow spin resonance.} \label{fig2}
\end{figure}

The resonance width can also be measured in the experiment by
locating the so called zero-crossing, i.e. the magnetic-field value
at which $a_{FB}$=0.  For example, we have done this for one of the
broadest resonances of our system by studying the efficiency of
sympathetic cooling of fermions close to the resonance. A mixture
composed by about 10$^6$ bosons and 10$^5$ fermions is loaded in a
crossed dipole trap at a temperature around 1~$\mu$K. We performed
this experiment using a Yb:YAG laser emitting at 1030~nm. We chose
relatively large beam waist radii of about 100~$\mu$m, in order to
have a relevant vertical sag of the rubidium cloud into the
non-parabolic part of the optical trap. This resulted in a more
efficient evaporation of Rb than of K when lowering the overall trap
depth via acusto-optic modulators. Fig.~\ref{fig2} shows the
temperature of the fermionic component after a 2.4~s evaporation in
vicinity of the broad Feshbach resonance in the ground state located
at 546.7 G. Thermalization to lower temperatures of spin-polarized
fermionic atoms can take place only via efficient fermion-boson
elastic collisions. Since the collisional cross-section at these low
temperature is well described by $\sigma$=$4\pi a_{FB}^2$, it
vanishes when $a_{FB}$ crosses zero. The zero-crossing location
extracted from a gaussian fit to the temperature of the fermionic
component is 543.4(1)~G. This gives a width $\Delta$=3.3(2)~G which
is in good agreement with the theoretical expectation of
$\Delta$=3~G. We have checked that such agreement between model and
experiment persists on a few other broad resonances of the mixture.

\section{Three-body losses at a Feshbach resonance}

The resonance in the ground state of the mixture around 546~G is in
principle the most appropriate to achieve a fine tuning of the
fermion-boson interaction. This is indeed one of the broadest
resonances, and moreover it can be expected to have the slowest
inelastic decay achievable for this mixture. Two-body inelastic
processes are indeed forbidden for the ground state, and only
three-body recombination is responsible for decay in proximity of
Feshbach resonances.

In a first series of experiments we have explored the main features
of three-body recombination in our system in the thermal regime, in
absence of possible interaction effects. At a boson-fermion Feshbach
resonance, three-body processes involving two bosons and one fermion
are the dominant decay channel, while those involving two fermions
and one boson are suppressed by the Pauli principle \cite{esry}.
During a three-body event, one boson and one fermion are associated
into a deeply bound dimer whose binding energy is shared as kinetic
energy by the dimer and the second boson. Conservation of energy and
momentum require this energy to be shared by the KRb dimer and the
Rb atom in a ratio of approximately 2:3. Typically this energy is at
least of the order of 100MHz$\times\hbar$, hence much larger than
the trap depth, and both atom and dimer are immediately lost.
\begin{figure}[thbp]
\includegraphics[width=\columnwidth,clip]{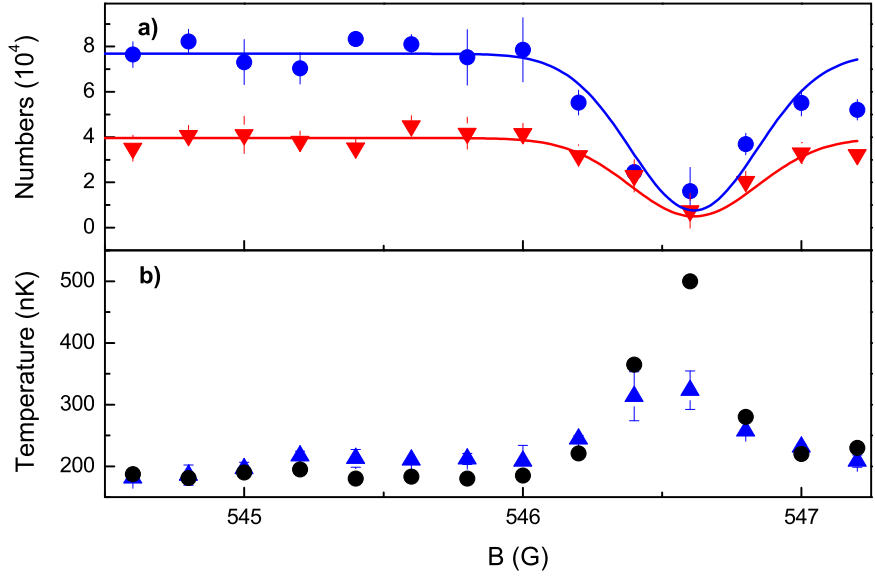}
\caption{Three-body recombination in a thermal fermion-boson mixture
after 100 ms of hold time at a Feshbach resonance. a) remaining atom
number for fermions (triangles) and bosons (circles). b) final
temperature for bosons measured in the experiment (triangles) and
calculated (circles).} \label{fig3}
\end{figure}
In the regime of very large scattering lengths the loss rate is
expected to depend on bosons and fermions density distributions
$n_B$ and $n_F$ and on the interspecies scattering length as
\begin{equation}
\Gamma_3=K_3\int n_B^2(\textbf{x})n_F(\textbf{x})d\textbf{x}\,,
\end{equation}
where $K_3\propto a_{FB}^4$ \cite{esry}. In general, these losses
result also in a heating of the remaining sample, since the
probability of finding two bosons and one fermion in the same
location is larger at the center of the distributions, where the
coldest atoms reside \cite{threebody}. The average energy per lost
particle can be computed as
\begin{equation}
E_l=\frac{\int U(\textbf{x})
n_B^2(\textbf{x})n_F(\textbf{x})d\textbf{x}}{\int{
n_B^2(\textbf{x})n_F(\textbf{x})d\textbf{x}}}\,,
\end{equation}
where $U$ is the potential energy of the trapped atoms. If the boson
and fermion distributions are equal, one finds the same heating rate
found for homonuclear bosonic systems: the energy per lost particle
is 0.5$k_BT$. This implies 3$k_BT$ excess energy for each loss
event, to be redistributed in the remaining sample.

In the experiment we have investigated the heating related to
three-body losses. For example, Fig.~\ref{fig3} shows the evolution
of atom numbers and temperature for a nondegenerate K-Rb sample that
was held for 100~ms at various magnetic fields around 546~G. We
clearly see a heating of the sample as the number of atoms drops
close to the resonance. Here we compare the temperature measured in
the experiment to the one calculated with an intuitive model. Such
model evaluates the time-evolution of the mixture as a series of
single three-body recombination events, each one followed by
rethermalization at a higher temperature due to the 3$k_BT$ excess
energy. Experiment and theory are in qualitative agreement, although
the latter slightly overestimates the heating.

\section{Tuning of the interaction in the quantum degenerate regime}

When the sample is cooled into the quantum degenerate regime in
proximity of the Feshbach resonance the properties of the system are
largely modified by the resonant interaction between the two
components. The Fermi-Bose interaction energy
\begin{equation}
U_{FB}(\textbf{x},a)=\frac{2\pi\hbar^2}{\mu}a_{FB}\int{n_B(\textbf{x},a_{FB})
n_F(\textbf{x},a_{FB})d\textbf{x}}\,,\label{eq2}
\end{equation}
can get comparable or even much larger than the trap potential, and
therefore determines the distributions of the two components in the
trap \cite{molmer}. In the case of $a_{FB}<$0 the large attractive
interaction can beat the natural repulsion within the Bose and Fermi
gases and lead to a collapse of the system. For large $a_{FB}>$0 the
two components can instead undergo phase separation.

Collapse of a Fermi-Bose mixture has already been observed in this
system as a sudden loss of a relevant fraction of the atoms when the
number of atoms in the condensate was increased along the
evaporation path \cite{hamburg}. The background K-Rb scattering length is indeed
negative and sufficiently large to reach the unstable region for
large atom numbers in tightly confining traps. No evidence of phase
separation has instead been reported in mixtures with positive
scattering length. The possibility of a rapid, fine tuning of the interaction at a
Feshbach resonance allows to access and characterize both regimes of
phase separation and collapse. In the experiment we have
investigated these regimes through a study of three-body losses in
the quantum degenerate regime. We indeed found that the loss
behavior can give strong indications on the overlap of the two
components in the trap.

To perform this experiment we produced a fully degenerate mixture by
evaporative cooling in the crossed dipole trap at 1030~nm. The trap
depth was lowered exponentially from 5~$\mu$K to 0.5~$\mu$K in
2.4~s, with a time constant of 1~s. The trap was then recompressed
to the full depth in 150~ms. The trap frequencies for Rb were
(120,92,126)~Hz and a factor about $\sqrt{(87/40)}$ larger for K.
The Fermi gas typically contains 5 10$^4$ atoms at a temperature
$T<0.3T_F$, where $T_F$=500~nK. The Bose-Einstein condensate
contains instead about 10$^5$ atoms and no thermal component is
discernible, i.e. the temperature is below 30~nK. The chemical
potential of the Bose gas is $\mu/k_B$=250~nK. Due to the different
energy scale, the linear sizes of the Fermi gas are almost twice
those of the Bose gas. The latter is totally contained in the volume
of the Fermi gas, and is shifted down by about 7~$\mu$m by the
larger gravitational force.

\begin{figure}[thbp]
\includegraphics[width=\columnwidth,clip]{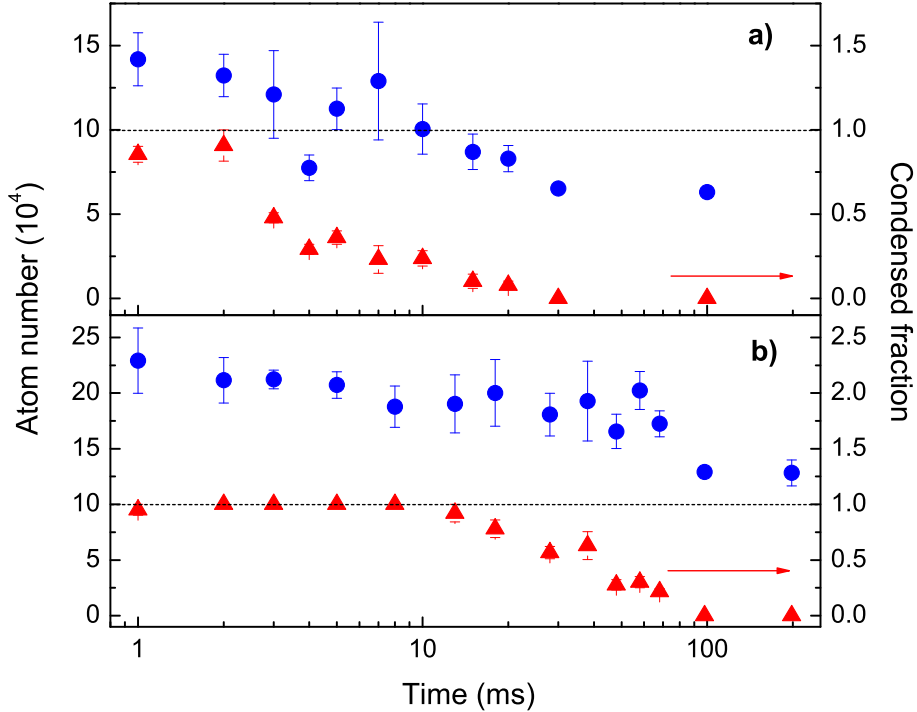}
\caption{Time-evolution of the boson number (triangles) and of the
condensed fraction (circles) for large positive and negative
scattering lengths: a) $a_{FB}$=-820$^{+40}_{-40}$; b)
$a_{FB}$=+740$^{+80}_{-70}$. In the first case the inelastic decay
and heating is sped up by interaction-induced collapse; in the
latter case it is slowed down by phase-separation.}\label{fig4}
\end{figure}

In a first experiment we have compared the time evolution of the
mixture for two values of the magnetic fields corresponding to
opposite values of the fermion-boson scattering length:
$a_{FB}\approx\pm$800~$a_0$. They were obtained by preparing the
mixture far from the resonance and then rapidly shifting the field
to the final value. For the positive $a_{FB}$ the mixture was
prepared at 539~G and then brought to 546.0~G, where the expectation
is $a_{FB}$=+740$^{+80}_{-70}$. It was instead prepared at 551~G for
negative $a$ and then brought to 547.6~G, where the expectation is
$a_{FB}$=-820$^{+40}_{-40}$. In such way crossing of the resonance
center during the rapid sweep was avoided. In Fig.~\ref{fig4} we
show the evolution of the condensed fraction of the Bose gas, which
is the most sensitive component of the mixture to excitations and
perturbations. On the negative $a_{FB}$ side we see a very rapid
depletion of the condensate on a timescale shorter than the trap
period, while for positive $a_{FB}$ the condensate remains stable
for a much longer time interval. At a longer time in both cases the
Bose gas is heated up into a pure thermal cloud. This however
happens already at about 20~ms for negative $a_{FB}$, and only at
about 100~ms for positive $a_{FB}$. During the whole time span, the
total atom number in the bosonic sample decreases by about 50\% in
the case of negative $a_{FB}$ and about 30\% for positive $a_{FB}$.
For the Fermi gas (not shown in the picture) we similarly observe
both atom loss and heating, which are larger for negative $a_{FB}$.

\begin{figure}[htbp]
\includegraphics[width=\columnwidth,clip]{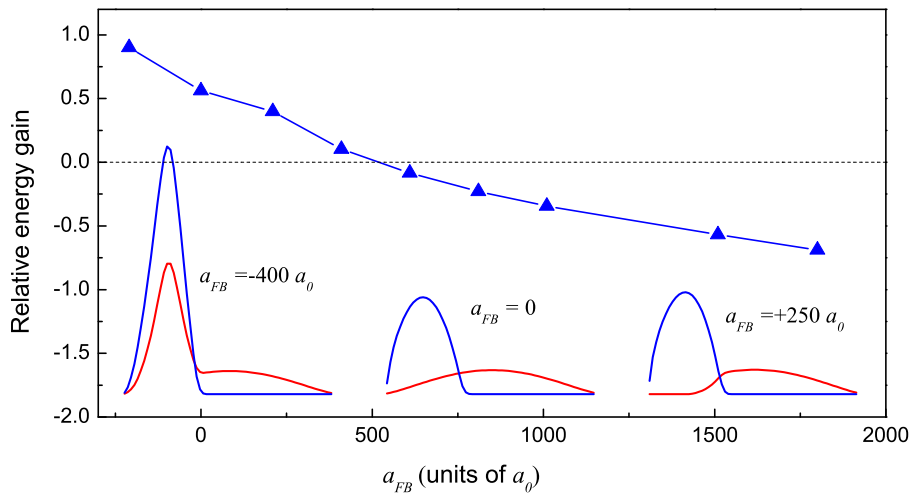}
\caption{Calculated relative energy gain
$(\overline{E}_B-E_l)/\overline{E}_B$ for each particle lost in a
three-body collision event for a K-Rb Fermi-Bose mixture with
variable interspecies interaction. For scattering lengths below
500~$a_0$ the system energy increases as a consequence of losses,
while above that value it decreases. This is a consequence of an
increased overlap in the first case, and of phase separation in the
second one, as shown by the calculated distributions along the
vertical direction of bosons and fermions at $T$=0. } \label{fig5}
\end{figure}

To understand the dramatically different behaviors observed for
positive and negative $a_{FB}$s we rely on a zero-temperature
mean-field model of our system \cite{modugnom}. We use it to study
the density distribution of the mixture in the two regimes, and to
predict the evolution of three-body losses. This model calculates
the local interaction energy and uses it as an additional effective
potential for both species to evaluate the distribution of the
mixture in the trap. This procedure is done recursively until the
true ground state of the system in presence of interaction is found.

For example, in Fig.~\ref{fig5} we plot a cut along the $z$
direction of the distribution of the two components when the number
of fermions and bosons is the same and equal to 5$\times$10$^4$, for
three different values of $a_{FB}$: 0, -400~$a_0$ and +250~$a_0$. In
the case of large negative $a_{FB}$ the density overlap of the two
components is clearly increased with respect to the non-interacting
case, while for large positive $a_{FB}$ the two components tend to
phase separate and it is strongly reduced. Note that the model would
not converge for $a_{FB}$=-800~$a_0$, since the system is collapsed,
i.e. it does not possess a stable ground state.

A reduced or increased overlap will obviously affect the three-body
loss rate $\Gamma_3$, which increases in the regime of collapse and
is reduced in the phase-separation regime. This will affects the
ratio between the mean energy of particles in the overlap region and
the mean energy of the whole system, which determines the heating
rate. We have used the model to calculate numerically the evolution
of the overlap integral, the mean energy per lost particle and the
mean energy per particle in the system, at $T$=0. For example, in
Fig.~\ref{fig5} we show the excess energy per lost particle,
normalized to the mean energy in the Bose condensate:
\begin{equation}
\frac{\overline{E}_B-E_l}{\overline{E}_B}\,,\,\,\,
\textrm{where:}\,\, \overline{E}_B= \frac{\int U(\textbf{x})
n_B(\textbf{x})d\textbf{x}}{\int n_B(\textbf{x})d\textbf{x}}\,.
\end{equation}
For $a_{FB}$=0 the model predicts a relative energy gain of about
0.5, close tho the classical value discussed above of 2/3. The
increase of heating on the $a_{FB}<0$ side and the corresponding
reduction on the $a_{FB}>0$ side are apparent. Actually, this model
indicates that three-body losses should eventually {\it cool down}
the system as $a_{FB}$ gets larger than 500$a_0$, where the relative
energy gain becomes negative.

Let us now interpret the behavior shown in Fig.~\ref{fig4} on the
basis of the model's predictions. For $a_{FB}\approx$-800$a_0$ the
system presumably starts a compression phase just after the
interaction energy is switched to a large and negative value. After
a quarter of the trap period, i.e. 2.5 ms, we have a maximum of
three-body loss rate and a large heating of the sample. The
condensate is therefore rapidly heated into a thermal cloud, and the
loss rate decreases because of the decreased density of the samples.
In the opposite case, $a_{FB}\approx$+800$a_0$, the system is in the
phase-separation regime. For the first 100 ms the condensate is not
heated up despite a 15$\%$ loss of atoms, indicating that
high-energy atoms are preferentially removed by the loss events.

Note that in order to have a more realistic, quantitative
description of the system's behavior, one would need a more complex
model which can also track the dynamics of the system at finite $T$.
For example, according to the simple model described above one would
expect to see a cooling of the system as the loss process goes on.
We think this is not observed because of the presence of the weakly
bound molecular state on this side of the resonance that is
responsible for the existence of the resonance itself. This can be
expected to result in the production of just moderately energetic
atom-molecule pairs, i.e. with a binding energy comparable to the
trap depth. These would have time to scatter with the remaining
atoms before leaving the trap, given the large collisional rate
expected in this regime (of the order of 1000 s$^{-1}$).

\begin{figure}[htbp]
\includegraphics[width=\columnwidth,clip]{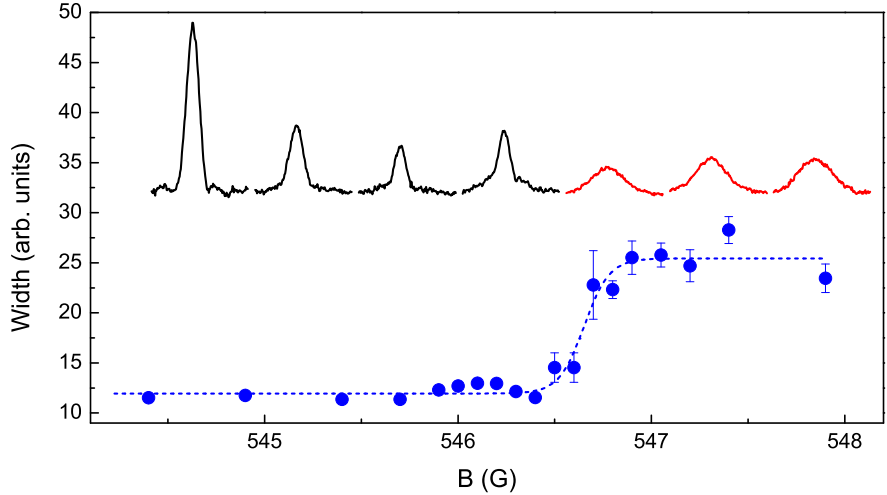}
\caption{Evolution of the Bose-Einstein condensate in the mixture
following a slow sweep across the Feshbach resonance from $a_{FB}>0$
to $a_{FB}<0$. Despite a strong reduction in the atom number due to
three-body losses, the system remains stable until the field crosses
the resonance center, where it is totally heated up into a thermal
sample. A phenomenological fit of the gaussian width of the
condensate with a Boltzmann growth function gives an accurate
estimation of the resonance center. } \label{fig6}
\end{figure}

The dramatically different behavior of losses on the two sides of
the resonance appears also in the measurements reported in
Fig.~\ref{fig6}. Here we report the evolution of the Bose-Einstein
condensate in the mixture following a sweep over the Feshbach
resonance. The field was increased in 50~ms from $B_i$=543.4~G to a
final field $B$ that was varied from 543.4~G to 548~G, and held
there for 10~ms. Inelastic losses start to deplete the system as the
sweep approaches the resonance center, but the condensate survives
as long as $a_{FB}>0$. It is instead very rapidly destroyed by
collapse when the sweep crosses the resonance center into the
$a_{FB}<0$ region. The evolution of the width of the gas after
ballistic expansion with $B$ can be used to find accurately the
resonance center, as shown in Fig.~\ref{fig6}b. A phenomenological
fit with a Boltzmann growth function gives $B_0$=546.65(20)~G, which
is in good agreement with the measured position of the loss maximum.

\begin{figure}[htbp]
\includegraphics[width=\columnwidth,clip]{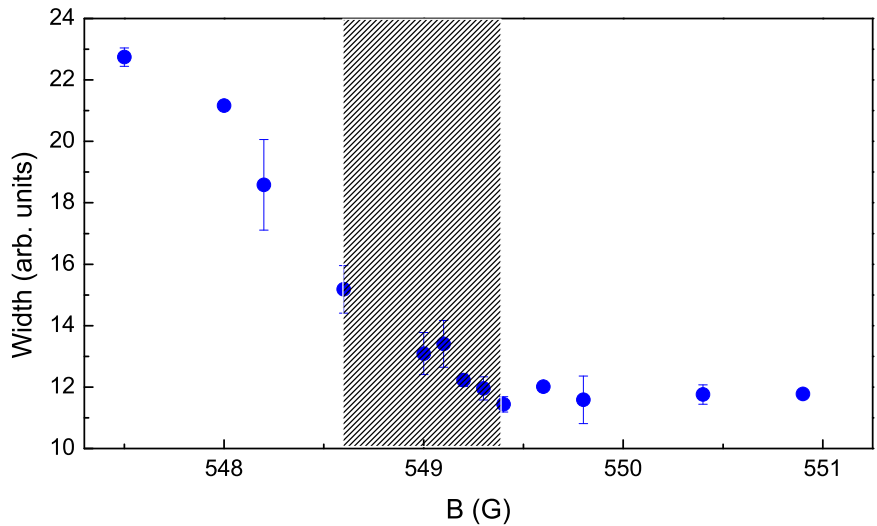}
\caption{Evolution of the width of the Bose-Einstein condensate in
the mixture following a sweep in the region of $a_{FB}<0$ close to a
Feshbach resonance. The highlighted area indicates the transition
magnetic-field region from a stable to an unstable (collapsed)
system.} \label{fig7}
\end{figure}

The possibility of a fine tuning of the interaction allows also to
explore the threshold for the collapse instability in the region
$a_{FB}<0$. According to our mean-field model, the system is
expected to remain stable as long as $a_{FB}>$-400~$a_0$, and to
collapse for larger negative scattering lengths. Fig.~\ref{fig7}
shows the evolution of the width of the Bose-Einstein condensate
after a sweep from above the resonance into the instability region.
The field was decreased in 50~ms from $B_i$=551~G to a lower final
field $B$, and held there for 20 ms. The width of the condensate
starts to increase around $B_f$=549.4~G until around $B_f$=548.6~G
the condensate is totally heated into a thermal sample as a
consequence of collapse. In the highlighted region we observe a
smooth transition between stable and unstable conditions: the
condensate is heated up by three-body recombination and shows shape
excitations due to the rapid change of the interaction energy . The
corresponding scattering length range is
$a_{FB}$=-600$\div$-350~$a_0$ in qualitative accordance with the
theory prediction.

Note that in this simple experiment it is difficult to make a more
precise determination of the threshold scattering length. It is
indeed difficult to separate an increase of the width due to a
complete collapse of the system from an increase due to ordinary
three-body losses in the regime of large attractive interactions or
simply a modified expansion of the Bose gas in presence of the Fermi
gas. Further investigation is needed in order to characterize the
whole phenomenology and possibly test mean-field models of this
system.

\section{Formation of dimers}

\begin{figure}[htbp]
\includegraphics[width=\columnwidth,clip]{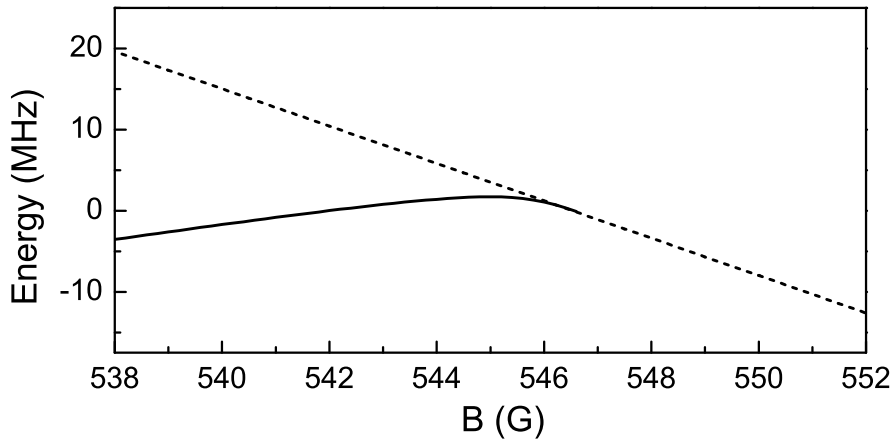}
\caption{Magnetic-field dependence of the atomic (dashed line) and
molecular (continuous line) states involved in the ground-state
 K-Rb Feshbach resonance. The atomic state is $|9/2, -9/2\rangle \otimes |1, 1\rangle$,
 while the molecular state is labeled as $F^{Rb}$=2, $\ell$=0,
 and has mixed singlet-triplet nature.}\label{fig8}
\end{figure}

An interspecies Feshbach resonance can also be exploited to
associate pairs of atoms into KRb dimers, using the same technique
that has proven successful in the case of homonuclear systems. The
idea is simple: since the resonance takes place in coincidence with
a crossing of an atomic and a molecular state, one can adiabatically
convert pairs of atoms into molecules with a magnetic-field sweep.
The magnetic-field dependence of the atomic and molecular state
involved in the resonance we studied are plotted in Fig.~\ref{fig8}.
The sweep needs to originate in the region $B>B_0$, where $E_a<E_m$,
and end on the other side. A maximum ramp speed can be evaluated
with a simple Landau-Zener model developed for homonuclear gases,
which describes the number of molecules as
\begin{equation}
N_{mol}=N_{max}(1-e^{-\delta_{LZ}})\;,\;\;\; {\rm where}\;\;\;
\delta_{LZ}=\alpha n\Delta/\dot{B},
\end{equation}
and $\alpha$=4.5(4)$\times$10$^4$~m$^2$s$^{-1}$ is an experimentally
determined coupling constant \cite{wieman}. The maximum conversion
is reached when $\delta_{LZ}$ gets larger than one, which in
our case corresponds to ramp speeds smaller than 50~G/ms.

\begin{figure}[htbp]
\includegraphics[width=\columnwidth,clip]{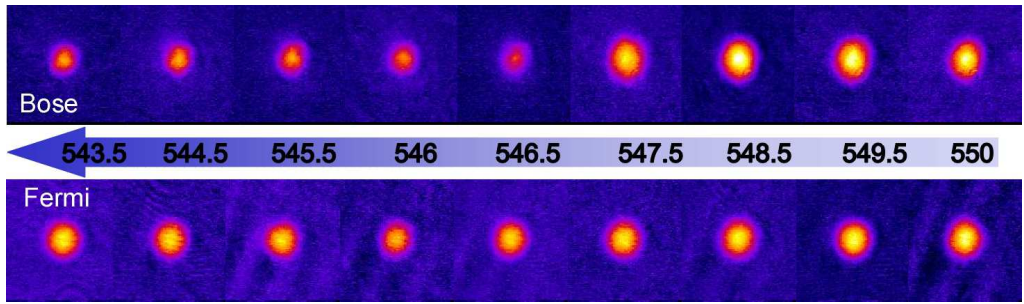}
\caption{Evolution of the atom numbers in the mixture during a
downward magnetic-field sweep at the ground-state Feshbach
resonance. The sudden decrease in atom number for both components
when the field crossed the resonance center at 546.6 indicates that
atom pairs are associated into KRb dimers.} \label{fig9}
\end{figure}

Fig.~\ref{fig9} shows a series of absorption images of the mixture
taken at various intermediate magnetic fields during a sweep over
the Feshbach resonances. The sweep was originating 4~G above the
resonance, and ended after about 5~ms at a variable magnetic field
across the resonance. The clouds were released from the optical trap
right at the end of the sweep, and the images were taken at zero
magnetic field, after an appropriate ballistic expansion. Note how
the number of atoms in both components drops as the field is brought
below 547~G. As in previous experiments, we interpret this reduction
in atom numbers as the result of molecule formation. The transition
energy of the molecules is indeed no longer resonant with the light
used to image the atoms, and molecules are therefore not detected.
It is important to note that the atoms are not lost because of
three-body recombination while sweeping over the resonance center,
where $a_{FB}\rightarrow\pm \infty$. Indeed, in that case one would
detect also a strong heating of the system, which is not apparent in
Fig.~\ref{fig9}.

Qualitative information on the process of molecule formation in this
system can be obtained through simple measurements. The maximum
conversion efficiency we were able to observe is about 30-40\%, that
is not obtained at the lowest temperatures achievable in the
experiment, but at temperatures around the condensation temperature
of bosons $T_c$. This can be qualitatively understood in terms of
the simple model and of the experiments on homonuclear systems
presented in \cite{wieman}. One expects to reach the maximum
conversion efficiency when the phase space overlap of the two
components is maximum. This is reached for $T$=0 in the homonuclear
Fermi or Bose cases, but not for a Fermi-Bose system, where the
spatial overlap of the two samples starts to decrease rapidly as
soon as $T$ gets smaller than $T_c$.

A crucial information for future experiments on such molecules is
obviously the stability of the molecular sample. This can be
investigated by reconverting the molecules into atoms via a backward
sweep across the resonance. We have been able to see molecules
converted back into atoms only under special conditions: very short
permanence on the molecular side of the resonance (shorter than
500~$\mu$s), and very low number of unpaired bosonic atoms. This
apparently indicates that the main decay channel of the molecular
sample is inelastic collisions with free bosonic atoms. Further
experiments in which one is able to remove the free atoms of both
species will be helpful to give a quantitative assessment of this
preliminary indication. A possible solution to the short lifetime of
the molecular sample is also the use of a tight 3D optical lattice.
This environment allows to prepare isolated K-Rb atomic pairs into
individual lattice sites, allowing for a loss-free production of KRb
molecules when sweeping over the resonance. As recently shown in
experiments \cite{ospelkaus}, this allows to have a lifetime
exceeding 100~ms, although this is obtained at the expenses of a
lower conversion efficiency.

\section{Outlook}

In this contribution we have discussed experiment on a K-Rb
Fermi-Bose mixture with tunable interaction at interspecies Feshbach
resonances. The capability of controlling the boson-fermion
scattering length to a high extent and the possibility of
associating pairs of atoms into molecules open few different
research directions.

The most important one is probably the study of ultracold dipolar
molecules. Weakly bound KRb molecules produced at Feshbach
resonances do not possess a relevant electric dipole moment, because
their range is so large that only a negligible distortion of the
electronic clouds is present. On the other hand, ground state
($X_1\Sigma^+$ ($\nu$=0)) KRb molecules are predicted to have a
dipole moment of the order of 1~D \cite{dipole}. These ground-state
molecules can in principle be produced by transferring "Feshbach
molecules" with optical Raman schemes \cite{raman} that are already
being developed \cite{storrs}. This is large enough to be used to
investigate the fundamental properties of ultracold and quantum
degenerate dipolar gases, and to test proposed schemes for quantum
computing applications, in which electric fields are used to address
and manipulate the particles \cite{demille}. The possibility of
controlling at will the Fermi-Bose interaction will also allow to
investigate novel quantum phases that have been proposed for
strongly correlated atomic systems in optical lattices
\cite{novelqp}. It can also allow to study in a more effective way
the phenomenology of disordered systems \cite{disorder}. Fine tuning
of $a_{FB}$ will also allow to test the predictions of present
theories of Fermi-Bose systems for collapse \cite{adikhari} and
phase-separation, collective excitations \cite{excitations}, effects beyond mean-field \cite{beyond}
and boson-induced superfluidity \cite{pairing}.

\acknowledgments I thank Massimo Inguscio, a continuous source of
ideas and encouragement, and Giacomo Roati, the driving force of the
Fermi-Bose experiment at LENS. I also thank the other people that
contributed to the work described in this lecture: Chiara D'Errico,
Francesca Ferlaino, Michele Modugno, Andrea Simoni, and Matteo
Zaccanti.  This work was supported by MIUR, by EU under contract
MEIF-CT-2004-009939, by Ente CRF, Firenze and by CNISM, Progetti di
Innesco 2005.


\begin{thebibliography}{99}

\bibitem{molecules} \BY{Jochim~S., Bartenstein~M., Altmeyer~A., Hendl~G., Riedl~S.,
Chin~C., Hecker Denschlag~J., \atque Grimm~R.}
\IN{Science}{302}{2003}{2101}; \BY{Greiner~M., Regal~C.~A., \atque
Jin~D.~S.} \IN{Nature}{426}{2003}{537}; \BY{Zwierlein~M.,
Stan~C.~A., Schunck~C.~H., Raupach~S.~M.~F., Gupta~S.,
Hadzibabic~Z., \atque Ketterle~W.} \IN{Phys. Rev.
Lett.}{91}{2003}{250401}.
\bibitem{fermi} \BY{Regal~C. A., Greiner~M., \atque Jin~D. S.}
\IN{Phys. Rev. Lett.}{92}{2004}{040403}; \BY{Chin~C.,
Bartenstein~M., Altmeyer~A., Riedl~S., Jochim~S., Hecker
Denschlag~J., \atque Grimm~R.} \IN{Science}{305}{2004}1128; \BY {
Bourdel~T., {\it et al.}}
 \IN{Phys. Rev. Lett.}{93}{2004}{050401}; \BY{Partridge~G. B., {\it et
al.}} \IN{Phys. Rev. Lett.}{95}{2005}{020404}; \BY{Zwierlein~M. W.,
Abo-Shaeer~J. R., Schirotzek~A., Schunck~C. H., \atque Ketterle~W.}
 \IN{Nature}{435}{2005}{1047}.
\bibitem{strongly} \BY{Winkler, K., {\it et al.}} \IN{Nature}{441}{2006}{853};
\BY{Stof\"{e}rle, T., {\it et al.}} \IN{Phys. Rev. Lett.}{96}{2006}{030401};
\BY{Chin~J.~K., {\it et al.}} \IN{Nature (London)}{443}{2006}{961}.
\bibitem{inouye} \BY{Inoyue~S., {\it et al.}} \IN{Nature
(London)}{392}{1998}{151}.
\bibitem{molmer} \BY{Molmer~K.} \IN{Phys. Rev. Lett.}{80}{1998}{1804}.
\bibitem{novelqp} \BY{Albus~A., Illuminati~F., \atque
Eisert~J.} \IN{Phys. Rev. A}{68}{2003}{023606};
\BY{B\"{u}chler~H.P., Blatter~G., \atque Zwerger~W.} \IN{Phys. Rev.
Lett.}{90}{2003}{130401};\BY{Lewenstein~M., Santos~L., Baranov~M.
A., \atque Fehrmann~H.} \IN{Phys. Rev. Lett.}{92}{2004}{050401}.
\bibitem{fmgases} \BY{Baranov~M. A., Marenko~M. S., Rychkov~V. S., \atque
Shlyapnikov~G.V.} \IN{Phys. Rev. A}{66}{2002}013606; \BY{Damski~B.,
{\it et al.}} \IN{Phys. Rev. Lett.}{90}{2003}{110401}.
\bibitem{zoller} \BY{Micheli~A., Brennen~G.K., Zoller~P.
}\IN{Nature Physics}{2}{2006}{341}.
\bibitem{mit} \BY{Stan~C. A., {\it et
al.}} \IN{Phys. Rev. Lett.}{93}{2004}{143001}.
\bibitem{ens} \BY{Zhang~J., {\it et al.}} in \TITLE{Proceedings of the XIX International Conference on
Atomic Physics}, edited by \NAME{L. G. Marcassa, V. S. Bagnato,
\atque K. Helmerson}(AIP, New York) 2005.
\bibitem{jila} \BY{Inouye~S., {\it et al.}} \IN{Phys. Rev. Lett.}{93}{2004}{183201}.
\bibitem{ferlaino} \BY{Ferlaino~F., D'Errico~C., Roati~G., Zaccanti~M., Inguscio~M., Modugno~G., \atque Simoni~A.} \IN{Phys. Rev. A}{73}{2006}{040702}.
\bibitem{zaccanti}\BY{Zaccanti~M., D'Errico~C., Ferlaino~F., Roati~G., Inguscio~M., \atque Modugno~G.} \IN{Phys. Rev. A}{73}{2006}{040702}.
\bibitem{ospelkausF}\BY{Ospelkaus~S., Ospelkaus~C., Humbert~L., Sengstock~K., \atque Bongs~K.}\IN{Phys. Rev.
Lett.}{97}{2006}{120403}.
\bibitem{roati}\BY{Roati~G., Riboli~F., Modugno~G. \atque Inguscio~M.} \IN{Phys. Rev.
Lett.}{89}{2002}{150403}.
\bibitem{ferrari} \BY{Ferrari~G., Jastrebski~W., Modugno~G., Roati~G., Simoni~A., \atque Inguscio~M.} \IN{Phys. Rev. Lett.}{89}{2002}{053202}.
\bibitem{sloshing}  \BY{Ferlaino~F., Brecha~R.~J., Hannaford~P., Riboli~F., Roati~G., Modugno~G.,
\atque Inguscio~M.} \IN{J. Opt. B: Quantum Semiclass.
Opt.}{5}{2003}{s3}.
\bibitem{simoni} \BY{Simoni~A., Ferlaino~F., Roati~G., Modugno~G., \atque
Inguscio~M.} \IN{Phys. Rev. Lett.}{90}{2003}{163202}.
\bibitem{science}\BY{Modugno~G., Ferrari~G., Roati~G., Brecha~R., \atque Inguscio~M.} \IN{Science}{294}{2001}{1320}.
\bibitem{munich} \BY{Marte~A.,{\it et al.}} \IN{Phys. Rev. Lett.}{89}{2002}{283202}.
\bibitem{derevianko} \BY{Derevianko~A., Babb~J. F., \atque Dalgarno~A.} \IN{Phys. Rev.
A}{63}{2001}{052704}.
\bibitem{goldwin} \BY{Goldwin~J., Inouye~S., Olsen~M. L., Newman~B., DePaola~B. D., \atque Jin~D. S.} \IN{Phys. Rev. A}{70}{2004}{021601}.
\bibitem{collapse} \BY{Modugno~G., Roati~G., Riboli~F., Ferlaino~F., Brecha~R.~J., \atque
Inguscio~M.} \IN{Science}{297}{2002}{2200}.
\bibitem{hamburg} \BY{Ospelkaus~C., Ospelkaus~S., Sengstock~K. \atque Bongs~K.} \IN{Phys. Rev. Lett.}{96}{2006}
{020401}.
\bibitem{zemke}\BY{Zemke~W.~T., Cot\'e~R., \atque Stwalley~W.~C.} \IN{Phys. Rev. A}{71}{2005}{062706}.
\bibitem{modugno}\BY{Modugno~G., Modugno~M., Riboli~F., Roati~G. \atque Inguscio~M.} \IN{Phys. Rev. Lett.}{89}{2002}{190404}.
\bibitem{minardi}\BY{Catani~J, Maioli~P., De Sarlo~L., Minardi~F., \atque Inguscio~M.} \IN{Phys. Rev. A}{73}{2006}{033415}.
\bibitem{cs} \BY{Leo~P., Williams~C.~J., \atque
Julienne~P.~S.} \IN{Phys. Rev. Lett.}{85}{2000}{2721}.
\bibitem{stuttgart} \BY{Werner~J., {\it et al.}} \IN{Phys.
Rev. Lett.}{94}{2005}{183201}.
\bibitem{bohnp} \BY{Ticknor~C., Regal~C. A., Jin~D. S., \atque Bohn~J. L.} \IN{Phys. Rev. A}{69}{2004}{042712}.
\bibitem{esry} \BY{D'Incao~J.~P. \atque Esry~B.~D.} \IN{Phys. Rev. A}{73}
{2006}{030702(R)}.
\bibitem{threebody} \BY{Weber~T., Herbig~J.,
Mark~M., N\"{a}gerl~H.-C, \atque Grimm~R.} \IN{Phys. Rev.
Lett.}{91}{2003}{123201}.
\bibitem{modugnom} \BY{Modugno~M., Ferlaino~F., Roati~G., Modugno~G., \atque Inguscio~M.} \IN{Phys. Rev. A }{68}{2003}{043626}.
\bibitem{wieman} \BY{Hodby~E., {\it et al.}} \IN{Phys. Rev. Lett.}{94}{2005}{120402}.
\bibitem{ospelkaus} \BY{Ospelkaus~C., {\it et al.}} \IN{Phys. Rev.
Lett.}{97}{2006}{120402}.
\bibitem{dipole}\BY{Kotochigova~S., Julienne~P.~S., \atque Tiesinga~E.} \IN{Phys. Rev.
A}{68}{2003}{022501}.
\bibitem{raman} \BY{Stwalley~W.~C.} \IN{Eur. Phys. J. D}{31}{2004}{221}.
\bibitem{storrs} \BY{Wang~D., Qi~J., Stone~M. F., Nikolayeva~O., Wang~H., Hattaway~B., Gensemer~S. D., Gould~P. L.,
Eyler~E. E., \atque Stwalley~W. C.} \IN{Phys. Rev.
Lett.}{93}{2005}{243005}; \BY{Sage~J. M., Sainis~S., Bergeman~T.,
\atque DeMille~D.} \IN{Phys. Rev. Lett.}{94}{2005}{203001}.
\bibitem{demille} \BY{DeMille~D.} \IN{Phys. Rev. Lett}{88}{2002}{067901}.
\bibitem{disorder} \BY{Gavish~U. \atque Castin~Y.} \IN{Phys. Rev. Lett.}{95}{2005}{020401};
\BY{G\"{u}nther~K., Sth\"{o}ferle~T., Moritz~H., K\"{o}hl~M., \atque Esslinger~T.}
\IN{Phys. Rev. Lett.}{96}{2006}{180402}; \BY{S. Ospelkaus, {\it et
al.}} \IN{Phys. Rev. Lett.}{96}{2006}{180403}.
\bibitem{adikhari} \BY{Adhikari~S. K.} \IN{Phys. Rev. A}{70}{2004}{043617}.
\bibitem{excitations} \BY{Liu~X.-J., Modugno~M. \atque Hu~H.} \IN{Phys. Rev.
A}{68}{2003}{053605}.
\bibitem{beyond} \BY{Albus~A. P., Illuminati~F., \atque Wilkens~M.} \IN{Phys. Rev. A}{67}{2003}{063606}.
\bibitem{pairing}  \BY{Heiselberg~H., Pethick~C. J., Smith~H., \atque Viverit~L.}
\IN{Phys. Rev. Lett.}{85}{2000}{2418}; \BY{Bijlsma~M. J.,
Heringa~A., \atque  Stoof~H.T.C.} \IN{Phys. Rev.
A}{61}{2000}{052601}; \BY{Viverit~L.} \IN{Phys. Rev.
A}{66}{2002}{023605}; \BY{Matera~F.} \IN{Phys. Rev.
A}{68}{2003}{043624}.
\end{thebibliography}
\end{document}